# Quantifying Human Mobility Perturbation and Resilience in Natural Disasters

Qi Wang[1], John E. Taylor[1]

**Human mobility is influenced by environmental change and natural disasters[1-3]. Researchers have used trip distance distribution, radius of gyration of movements, and individuals' visited locations to understand and capture human mobility patterns and trajectories[4-7]. However, our knowledge of human movements during natural disasters is limited owing to both a lack of empirical data and the low precision of available data. Here, we studied human mobility using high-resolution movement data from individuals in New York City during and for several days after Hurricane Sandy in 2012. We found the human movements followed truncated power-law distributions during and after Hurricane Sandy, although the $\beta$ value was noticeably larger during the first 24 hours after the storm struck. Also, we examined two parameters: the center of mass and the radius of gyration of each individual's movements. We found that their values during perturbation states and steady states are highly correlated, suggesting human mobility data obtained in steady states can possibly predict the perturbation state. Our results demonstrate that human movement trajectories experienced significant perturbations during hurricanes, but also exhibited high resilience. We expect the study will stimulate future research on the perturbation and inherent resilience of human mobility under the influence of natural disasters. For example, mobility patterns in coastal urban areas could be examined as tropical cyclones approach, gain or dissipate in strength, and as the path of the storm changes. Understanding nuances of human mobility under the influence of disasters will**

---

[1] Civil Engineering Network Dynamics Lab, Charles E. Via, Jr. Department of Civil and Environmental Engineering, Virginia Tech, Blacksburg, Virginia 24061, USA.

**enable more effective evacuation, emergency response planning and development of strategies and policies to reduce fatality, injury, and economic loss.**

Recently, one of the most powerful tropical storms on record—Typhoon Haiyan—struck the Philippines resulting in over 6,000 deaths, 4.1 million people displaced and 14.1 million affected[8]. According to the United Nations Office for Disaster Risk Reduction, over the period from 2000 through 2012 natural disasters have been responsible for killing 1.2 million people and affected the lives of 2.9 billion people[9]. This tragic loss of life and human suffering calls for a better understanding of human mobility during typhoons and hurricanes to aid evacuation, emergency response and immediate post-disaster relief[4,10,11]. Improved knowledge about human mobility can be used to reduce traffic jams[12], establish temporary communication networks[13,14], and deliver critical information to reduce injuries, fatalities, and economic loss[15,16].

Recent research has improved our understanding of general human mobility patterns but unfortunately we know comparatively little about human movements during disasters. Research has shown that human movements follow power-law distributions with $\beta$ values ranging from 1.59 to 1.75[4,5]. Individual movement trajectories exhibited similar shapes after being rescaled by the radius of gyration[4]. But, movement trajectories demonstrate uneven visitation frequency of different locations, repeatedly returning to certain locations while being less likely to visit new ones[6,7]. At the city scale, movements follow similar distributions in different urban areas[17] and exhibit characteristics of periodicity to return to primary locations and unusual bursts during special events[18]. Also, human movements have been shown to follow highly efficient trajectory configurations during their daily movements[19]. While human mobility studies have improved our knowledge about general mobility patterns, it is intuitive to assume that disasters may perturb such routine movements[1,2,20] and even cause population migration[3]. Therefore, human mobility trajectories during disasters would deviate from steady states. Research on the topic of human mobility under the influence of disasters is limited perhaps owing to the inherent unpredictability of disasters and resulting data scarcity. Moreover, much empirically grounded human mobility research utilizes mobile phones to track human mobility[4,6,7,19,21]. The data precision of these studies is limited to the coverage area of each mobile phone tower, which is typically around 3km$^2$. While such precision has been instrumental in developing an understanding of general patterns of human mobility over larger scales (e.g., a state or a country), it may lack the necessary precision to capture mobility changes caused by disasters and other extreme events that unfold at smaller scales (e.g., a building or a city). The necessity of high-resolution data cannot be ignored because many of those caught in the storm surge in Tacloban during Typhoon Haiyan—which was responsible for most fatalities—may have been spared if they had evacuated a short distance to higher ground.

We study the perturbation and resilience of human mobility patterns during and after Hurricane Sandy—one of the largest tropical storms recorded in the Atlantic Ocean that affected millions of people[22,23]—which struck the northeastern seaboard of the United States leading to significant injury and loss of human life in October 2012. The study used high-resolution data from Twitter which were collected from 4pm Oct. 29, 2012—the day Hurricane Sandy landed in New York City—through 4pm Nov. 10, 2012. During the 12-day data collection period, a total of 702,188 tweets were collected from a total of 53,934 distinct individuals. We located each user using geolocation information attached to each tweet which contains a longitude and latitude geographical coordinate. The coordinates are high resolution with precision to approximately 0.001 meter and accuracy to 3 meters.

After mapping each recorded location during every 24-hour period, we observed that movement locations covered nearly the entire mapped area and showed similar geographical distribution to 24-hour periods soon after the hurricane (Fig. 1a, 1c, and 1e). This observation suggests that New York City residents were relatively resilient in terms of human mobility during Hurricane Sandy. While such resilience could be vital for the city's post-disaster response and recovery, it may also be life threatening during an extreme event such as a hurricane. Overlapping the location and movement data with the mandated evacuation areas reveals that human activities were still observed in evacuation zones although people were ordered to evacuate, (Fig. 1a and 1b). Regrettably, several fatalities occurred in these evacuation zones.

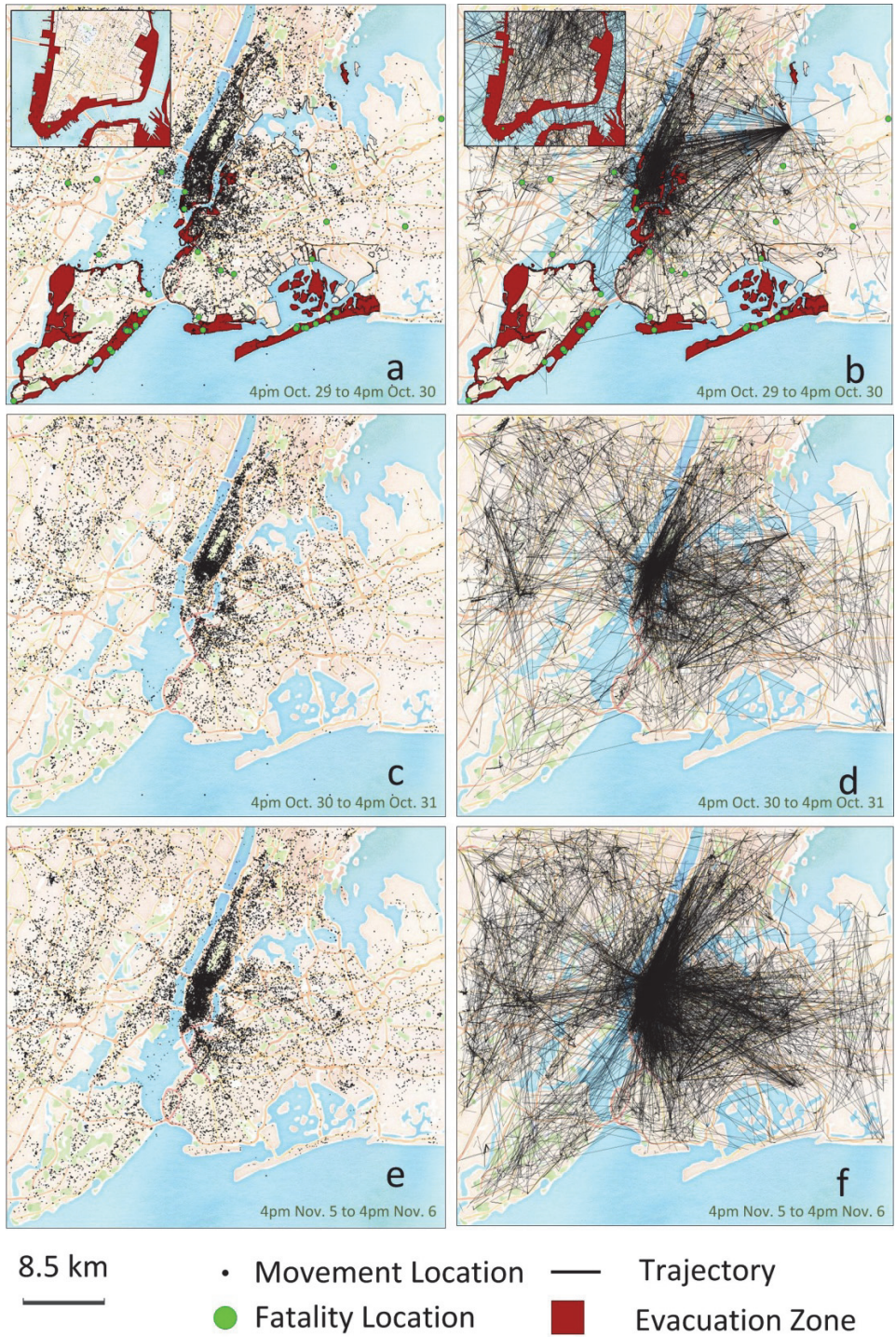

**Figure 1 | Geographical distribution of visited locations and movement trajectories over 24-hour periods. a**, **c**, and **e**, locations visited by Twitter users. **b**, **d**, and **f**, movement trajectories of Twitter users. The insets in **a** and **b** show an enlarged map of the lower Manhattan area. Red areas indicate the evacuated zones enforced by New York City government, though some of the areas were still active with human activity/mobility in this 24-hour period. The green nodes indicate the locations where fatalities occurred.

To discover whether human movements were disrupted, we also mapped each individual's unique mobility trajectory during each 24-hour period. We found the movement trajectories in the post-disaster period exhibited similar patterns (Fig. 1d and 1f), radiating from Manhattan to other areas, with the exception of the trajectories in the first 24-hour period. To test if the observed difference was mathematically significant, we posed the following hypothesis:

**Hypothesis 1:** During Hurricane Sandy, the human mobility distribution was significantly perturbed.

To explore this hypothesis, we calculated all the distances between consecutive locations from each distinct user. We found that while short-distance trips ($\Delta r$<1km) increased during the first 24-hour period, longer-distance travels significantly decreased (Fig. 2a). Also, the total displacements ($\Delta r$) for each 24-hour period followed a truncated power-law distribution, though with different values (Fig. 2b). During the first 24-hour period during which Hurricane Sandy struck New York City, the $\beta$ value of the distribution was 1.73, the exponential cutoff value $\lambda$ was $2.70 \times 10^{-5}$, and the minimum fitting value $\kappa$ was 591 m. During the 11 other 24-hour data collection periods, the $\beta$ value was $1.19 \pm 0.06$, the $\lambda$ value was $(6.13 \pm 0.69) \times 10^{-5}$, and the cutoff value $\kappa$ was 4 m. The change of the $\beta$ values is primarily the result of the larger optimized cut-off value of the first 24-hour displacement data. The reduction of longer-distance travels and the substantial difference between $\beta$ and $\kappa$ allowed us to reject the null hypothesis and confirm that a perturbation of human mobility occurred during Hurricane Sandy. The differences between the calculated $\beta$ values here and the values reported in other studies[4,5] perhaps are owing to the fact that the values in our study were derived from higher precision location data in a more tightly constrained geographical area. It is worth noting, however, that the truncated power-law distribution discovered here is similar to the human mobility patterns identified in those studies.

The results confirm that Hurricane Sandy perturbed human mobility patterns. However, the results reveal other important insights regarding human mobility perturbation during Hurricane Sandy in New York City. First, the distribution of the perturbation bears strong similarity to daily movement distributions in the following 11-day period. Second, the significance of the perturbation appears to have lasted less than 24 hours as the aggregate mobility distribution returned to a steady state on Oct. 30, 2012. Human mobility in New York City was resilient to the effects of Hurricane Sandy. This resilience in human mobility was the case even as most public transportation only started to resume partial or full schedule service 36 to 72 hours after Hurricane Sandy struck[24] and over 1 million people in the city were still without power until 2pm Nov. 2[25].

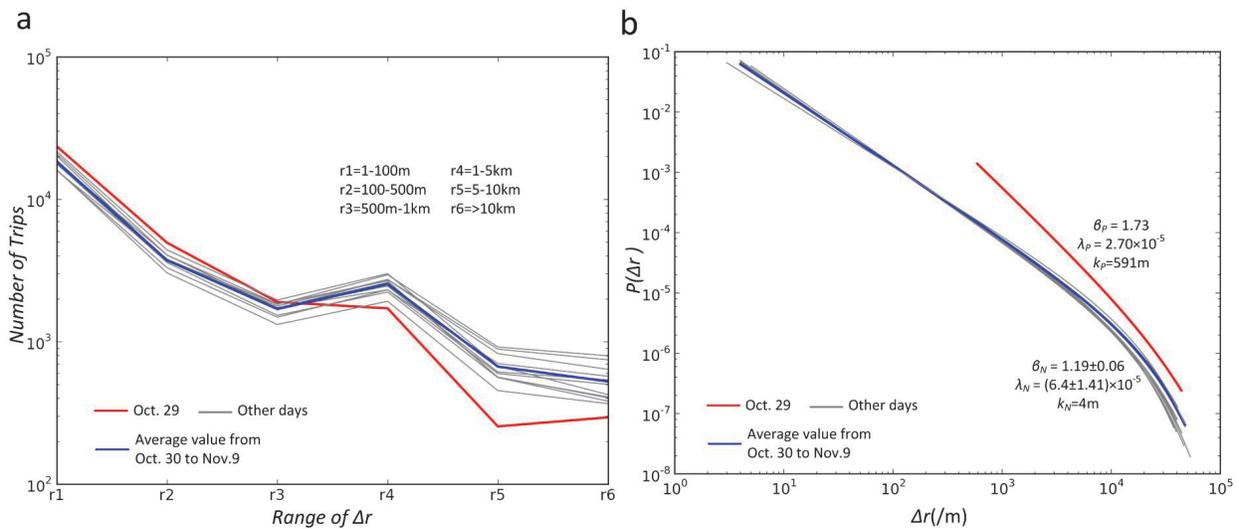

**Figure 2 | Human Mobility Perturbation. a**, Statistical distribution of displacements for each 24-hour period. Each line represents the change of numbers of trips within different ranges of displacements. **b**, Displacement distribution for each 24-hour period. Each line represents a probability density function $P(\Delta r)$. All plotted distributions followed a truncated power-law distribution (Supplementary Table 3).

While Fig. 2a and 2b show a clear deviation between the mobility distribution of the first 24-hour period and the distributions of the other 24-hour periods, demonstrating that Hurricane Sandy perturbed human mobility in New York City, it raises new questions: how did people change their movements and do these changes correlate to movements in non-perturbed states? To explore these questions, we mapped each individual's movement trajectory over each 24-hour period to examine perturbations at the individual level. Individuals generally exhibited two perturbation trends: shifting their movements to a different place or places, and changing their travel distances. To quantify the observed perturbation, we calculated the center of mass ($\vec{r}_{CM}$) and the radius of gyration ($r_g$) for each person's trajectory during the 24-hour perturbation state (*P*) and the relatively steady state that resulted in the ensuing 24-hour periods (*N*). The perturbation can be defined by two parameters: the shifting distance of the center of mass $\Delta d_{CM} = \vec{r}_{CM}^{P} - \vec{r}_{CM}^{N}$, and the radius of gyration in perturbation state $r_g^P$. To examine whether the values of these parameters can be projected using the values from non-perturbed states or whether disasters make human movements chaotic, we tested two additional hypotheses:

**Hypothesis 2a:** $\Delta d_{CM}$ correlates with $r_g^N$.

**Hypothesis 2b:** $r_g^P$ correlates with $r_g^N$.

We first analyzed $\Delta d_{CM}$ and found that the values followed a stretched exponential distribution (Supplementary Fig. 2). We also found the values strongly correlate with $r_g^N$ (correlation coefficient = 0.25, $p<0.001$). Larger values of $r_g^N$ can result in larger shifting distances in the center of movements for individuals. The shifting distance $\Delta d_{CM}$ follows $\Delta d_{CM} = 204 r_g^{N^{0.32}}$ (Fig. 3a). Therefore, we can reject the null hypothesis for Hypothesis 2a, finding support that $r_g^N$ can possibly be used to predict the shift of the center of movements. The result is interesting but perhaps not surprising. It implies that a person needs to move to a safer place during a hurricane in order to sustain their mobility, albeit over shorter distances.

We then examined whether a correlation exists between $r_g^P$ and $r_g^N$. We found that both $r_g^P$ and $r_g^N$ follow truncated power-law distributions (Supplementary Fig. 3). The value of $r_g^N$ was 1.24, approximating the value reported in a previous study[4] (~1.20). However, the $\beta$ value of $r_g^P$ was found to be $1+4.8\times 10^{-8}$, which suggests a perturbation in the radius of gyrations in movement trajectory. Pairing $r_g^P$ and $r_g^N$, we found their correlation can be captured with the function $r_g^P = 166 r_g^{N^{0.22}}$ (Fig. 3b). The correlation between the two parameters resulted in a correlation coefficient of 0.23 ($p<0.001$), and therefore, the null hypothesis for Hypotheses 2b can be rejected. This implies that $r_g^N$ can be used to predict $r_g^P$.

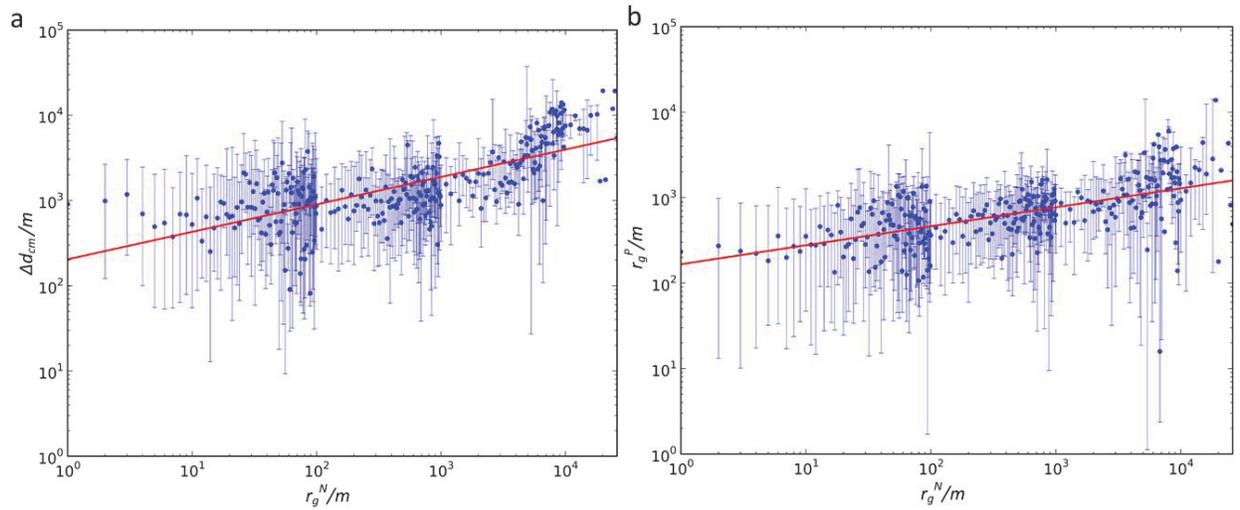

**Figure 3 | Relation between perturbation states and steady states. a**, $\Delta d_{CM}$ versus $r_g^N$. The blue node indicates the median value of $\Delta d_{CM}$ at different values of $r_g^N$, and the bars represent the ranges. The correlation coefficient between the two parameters is 0.25 ($p<0.001$). The red line is the fitted function of the correlation where $\Delta d_{CM} = 204 r_g^{N^{0.32}}$ (/m). **b,** $r_g^P$ versus $r_g^N$. The notations are the same as in **a**. The correlation coefficient between these two parameters is 0.23 ($p<0.001$) with fitted function $r_g^P = 166 r_g^{N^{0.22}}$ (/m).

Our results demonstrate that, although Hurricane Sandy did impact the mobility patterns of individuals in New York City, the perturbation was surprisingly brief and the mobility patterns encouragingly resilient. This resilience occurred even in the large-scale absence of mobility infrastructure. While human movements followed truncated power-law distributions in perturbed and non-perturbed states, their mobility-related behaviors did exhibit important changes. That is, people changed the center of their movements and their travel distances as they sought shelter.

Our finding that steady state shifts in the center of mass ($\Delta d_{CM}$) and the radius of gyration ($r_g^N$) of daily individual trajectories correlate with perturbed states is particularly interesting. Human mobility appears to possess an inherent resilience—even in perturbed states—such that movement deviations, in aggregate, follow predictable patterns in hurricanes. Therefore, it may be possible to use human mobility data collected in steady states to predict perturbation states during extreme events and, as a result, develop strategies to improve evacuation effectiveness and speed critical disaster response to minimize loss of life and human suffering.

**References:**

1. Horanont, T., Phithakkitnukoon, S., Leong, T. W., Sekimoto, Y. & Shibasaki, R. Weather Effects on the Patterns of People's Everyday Activities: A Study Using GPS Traces of Mobile Phone Users. *PloS one* **8**, e81153. (2013).
2. Bagrow, J. P., Wang, D. & Barabási, A.-L. Collective response of human populations to large-scale emergencies. *PloS one* **6**, e17680 (2011).
3. Bengtsson, L., Lu, X., Thorson, A., Garfield, R. & von Schreeb, J. Improved response to disasters and outbreaks by tracking population movements with mobile phone network data: a post-earthquake geospatial study in Haiti. *PLoS medicine* **8**, e1001083 (2011).
4. González, M. C., Hidalgo, C. A. & Barabási, A.-L. Understanding individual human mobility patterns. *Nature* **453**, 779-782 (2008).
5. Brockmann, D., Hufnagel, L. & Geisel, T. The scaling laws of human travel. *Nature* **439**, 462-465 (2006).
6. Song, C., Koren, T., Wang, P. & Barabási, A.-L. Modelling the scaling properties of human mobility. *Nat Phys* **6**, 818-823, (2010).
7. Song, C., Qu, Z., Blumm, N. & Barabási, A.-L. Limits of predictability in human mobility. *Science* **327**, 1018-1021 (2010).
8. CWS(ASIA/PACIFIC). Typhoon Haiyan/Yolanda Philippines: Situation Report 19 – December 19[th], 2013. (2013). Available at: http://reliefweb.int/sites/reliefweb.int/files/resources/2013.12.19-Haiyan-Typhoon-Philippines-Situation-Report-19.pdf. Web: Feb 9, 2014.
9. UNISDR. Disasters Impacts 2000-2012. (2013). Available at: http://www.preventionweb.net/files/31737_20130312disaster20002012copy.pdf. Web: Feb. 9, 2014.

| | |
|---|---|
| 10 | Candia, J. *et al.* Uncovering individual and collective human dynamics from mobile phone records. *Journal of Physics A: Mathematical and Theoretical* **41**, 224015 (2008). |
| 11 | Adger, W. N., Hughes, T. P., Folke, C., Carpenter, S. R. & Rockström, J. Social-ecological resilience to coastal disasters. *Science* **309**, 1036-1039 (2005). |
| 12 | Pan, X., Han, C. S., Dauber, K. & Law, K. H. A multi-agent based framework for the simulation of human and social behaviors during emergency evacuations. *Ai & Society* **22**, 113-132 (2007). |
| 13 | Chaintreau, A. *et al.* Impact of human mobility on opportunistic forwarding algorithms. *Mobile Computing, IEEE Transactions on* **6**, 606-620 (2007). |
| 14 | Feeley, M., Hutchinson, N. & Ray, S. in *Ad-Hoc, Mobile, and Wireless Networks* 324-329 (Springer-Verlag Berlin Heidelberg,German), (2004). |
| 15 | Vespignani, A. Complex networks: The fragility of interdependency. *Nature* **464**, 984-985 (2010). |
| 16 | Kleinberg, J. Computing: The wireless epidemic. *Nature* **449**, 287-288 (2007). |
| 17 | Noulas, A., Scellato, S., Lambiotte, R., Pontil, M. & Mascolo, C. A tale of many cities: universal patterns in human urban mobility. *PloS one* **7**, e37027 (2012). |
| 18 | Liang, X., Zheng, X., Lv, W., Zhu, T. & Xu, K. The scaling of human mobility by taxis is exponential. *Physica A: Statistical Mechanics and its Applications* 391, 2135-2144 (2012). |
| 19 | Schneider, C. M., Belik, V., Couronné, T., Smoreda, Z. & González, M. C. Unravelling daily human mobility motifs. Journal of The Royal Society Interface 10 (2013). |
| 20 | Morrow-Jone, H. A. & Morrow-Jone, C. R. Mobility due to natural disaster: Theoretical considerations and preliminary analyses. Disasters 15, 126-132 (1991). |
| 21 | Wesolowski, A. et al. Quantifying the impact of human mobility on malaria. Science 338, 267-270 (2012). |
| 22 | Ferris, E., Petz, D. & Stark, C. The Year of Recurring Disaster: A Review of Natural Disasters in 2012. (The Brokings Institution) (2013). |
| 23 | OCHA. The Caribbean: Hurricane Sandy Situation Report No. 2 (as of 19 November 2012). (2012). |
| 24 | Kaufman, S., Qing, C., Levenson, N. & Hanson, M. Transportation During and After Hurricane Sandy. (Rudin Center for Transportation NYU Wagner Graduate School of Public Service, 2012)(2012). |
| 25 | McGeeham, P. Wait for power may linger for some in *The New York Times* Nov 2, 2012: A20 (2012). |

**Acknowledgments** This study is supported by the National Science Foundation under Grant No. **1142379**. Any opinions, findings, and conclusions or recommendations expressed in this material are those of the authors and do not necessarily reflect the views of the National Science Foundation.

**Author Information** Human mobility data were collected using the open API provided by Twitter. Due to Twitter's policy, the data are not available for distribution. The maps used in this paper are developed by OpenStreetMap under CC BY SA, and tiled by Stamen Design under CC BY 3.0. Correspondence and requests for material should be addressed to J.E.T (jet@vt.edu).